\documentclass[aps,prl,preprint]{revtex4-1}

\usepackage{xcolor}

\usepackage{graphicx} 

\usepackage{epstopdf}

\usepackage{bm}

\begin{document}

\title{\LARGE \bf
Mapping different skyrmion phases in double wedges of Ta/FeCoB(t$_{FeCoB}$)/Ta(t$_{Ta}$)Ox }
\author{Titiksha Srivastava}
\altaffiliation[Now at ] {UMPhy, CNRS-Thales, Palaiseau, France}%

\affiliation{ Univ. Grenoble Alpes, CEA, CNRS, Grenoble INP, IRIG, SPINTEC, F-38000 Grenoble, France}%
\author {Willy Lim}
\affiliation{ Univ. Grenoble Alpes, CEA, CNRS, Grenoble INP, IRIG, SPINTEC, F-38000 Grenoble, France}
\author{Isabelle Joumard}
\affiliation{ Univ. Grenoble Alpes, CEA, CNRS, Grenoble INP, IRIG, SPINTEC, F-38000 Grenoble, France}
\author{St\'ephane Auffret}
\affiliation{ Univ. Grenoble Alpes, CEA, CNRS, Grenoble INP, IRIG, SPINTEC, F-38000 Grenoble, France}
\author{Claire Baraduc}
\affiliation{ Univ. Grenoble Alpes, CEA, CNRS, Grenoble INP, IRIG, SPINTEC, F-38000 Grenoble, France}
\author{H\'el\`ene B\'ea}
\affiliation{ Univ. Grenoble Alpes, CEA, CNRS, Grenoble INP, IRIG, SPINTEC, F-38000 Grenoble, France}

\begin{abstract}
Skyrmions are chiral magnetic textures that have immense potential for applications in spintronic devices. However, their formation is quite challenging and necessitates a subtle balance of the magnetic interactions at play. Here, we study Ta/FeCoB/TaOx trilayer using crossed double wedges i.e. thickness gradients of FeCoB and of top Ta, which is subsequently oxidized, leading to an oxidation gradient. This enabled us to observe micron-sized skyrmions in the vicinity of different transition regions of the sample: from perpendicular magnetic anisotropy to paramagnetic phase and also from perpendicular to in-plane magnetic anisotropy. These observations can be explained by the isolated skyrmion model taking into account the different energy contributions at play namely anisotropy, exchange, Dzyaloshinskii-Moriya, dipolar and Zeeman. We also qualitatively compare the current-induced motion of skyrmions obtained in different transition regions. Our study not only provides an effective means to form skyrmions by tuning the interfacial magnetic properties  but also highlights the differences pertaining to the skyrmions observed in different transition zones, which is extremely crucial for any envisaged application. 
\end{abstract}

\maketitle


The discovery of topologically non-trivial chiral magnetic configurations called skyrmions \cite{Bogdanov,Ivanov} ignited a new possibility of ultra-fast and robust magnetic memories and logic devices. One key ingredient for skyrmion homochirality lies in Dzyaloshinskii-Moriya interaction (DMI) which is an antisymmetric exchange interaction induced by spin orbit coupling in the presence of broken inversion symmetry. Skyrmions were initially observed in bulk non centro-symmetric crystals like MnSi and FeGe at low temperature \cite{Yu,Muhlbauer} and have more recently been identified in thin trilayer systems at room temperature consisting of a heavy metal (HM), a ferromagnet (FM) and an insulator (usually metal oxides MOx) like Ta/FeCoB/TaOx  \cite{Jiang}, Pt/CoFeB/MgO \cite{Woo} or Pt/Co/MgO \cite{Boulle}, in HM/FM/HM multilayers (Pt/Co/Os/Pt \cite{Tolley}) or repetitions of such multilayers \cite{Woo, Luchaire}. In HM/FM/MOx trilayer systems, an interfacial DMI arises either at the interfaces between FM and  HM (Fert-Levy mechanism \cite{Fert}) and/or between FM and  MOx (Rashba mechanism \cite{Kim,Shanavas,Yang}). In these trilayer structures, FM, HM and oxide film thicknesses and also interfacial properties can be finely and independently adjusted: it is a way of tuning magnetic properties for stable skyrmion generation and manipulation and is well suited to bring skyrmions closer to technological integration.

The total magnetic energy minimization required to stabilize skyrmions is composed of competing interactions: Heisenberg exchange, Zeeman, dipolar, anisotropy (magnetocrystalline and interface contributions) and DMI. These different contributions of volume/interface origin evolve differently with film thicknesses and interface qualities so that a complex magnetic configuration diagram can be constructed \cite{ABM,Buttner}. Therefore, obtaining magnetic skyrmions at room temperature is not straightforward. Most of the methods rely on geometrical patterning and application of current to nucleate and stabilize them \cite{Jiang, Boulle, Juge,Legrand}.

Another way is to nucleate them in an extended uniform film where the energy balance favors demagnetized domain configuration at room temperature. Indeed, when magnetic parameters allow Perpendicular Magnetic Anisotropy (PMA) with low domain wall energy, it may lead to spontaneous nucleation of many up-down domains (stripes) at room temperature to decrease dipolar energy costs. Upon introducing a small out-of-plane magnetic field, these stripes collapse into N\'eel type-skyrmionic bubbles \cite{Titiksha,Marine} (if DMI is large  enough \cite{Thiaville}) with their core magnetization pointing opposite to the direction of the applied magnetic field. 
Several ways can be employed to maneuver the volume and interface magnetic properties contributions to stabilize skyrmion phases. Some studies have shown interface engineering method using thin insertions of heavy metal in between the ferromagnet and insulator \cite{Yu2016}. Some have demonstrated that the oxidation state of the insulator on the top, which is generally a metal oxide, also affects the interfacial magnetic properties and thus perpendicular magnetic anisotropy  \cite{Zhu,Manchon} and DMI \cite{Belabbes,deSouza}. In Ta/FeCoB/TaOx trilayers, skyrmions have been observed at room temperature and have shown promising results towards current driven motion via spin-orbit torques and also expected voltage driven chirality switching \cite{Jiang,Titiksha}.

In this letter, we therefore investigate in detail stable energy landscapes for skyrmion phases in Ta/FeCoB/TaOx trilayer by studying the magnetic properties as a function of ferromagnetic thickness and insulating layer oxidation state. This skyrmion preparation method from thermally demagnetized stripes thus requires a fine tuning of domain wall energy given by $\sigma_W=4\sqrt{AK}-\pi \mid D\mid$ \cite{Heide} with $K$ the effective anisotropy ($K=K_s/t -1/2 \mu_0 M_s^2$ where $K_s$ is the surface anisotropy and $t$ is the effective ferromagnetic film thickness),  $A$ the exchange stiffness and $D$ the DMI energy (in $J/m^2$). Low domain wall energy can be achieved either by reducing anisotropy $K$ or for finite $K$ adding DMI that reduces domain wall cost for a given chirality.
We therefore used crossed double wedges of FeCoB and top Ta that is subsequently oxidized to locate different zones where skyrmions are energetically stable under a very small applied magnetic field: they appear in the vicinity of transitions from PMA to paramagnetic  state and from PMA to In-Plane Anisotropy (IPA). We also demonstrate a comparison of the current induced motion of skyrmions located in different zones of the double wedge.


\begin{figure}
	\centering
	\includegraphics[width=0.98\columnwidth]{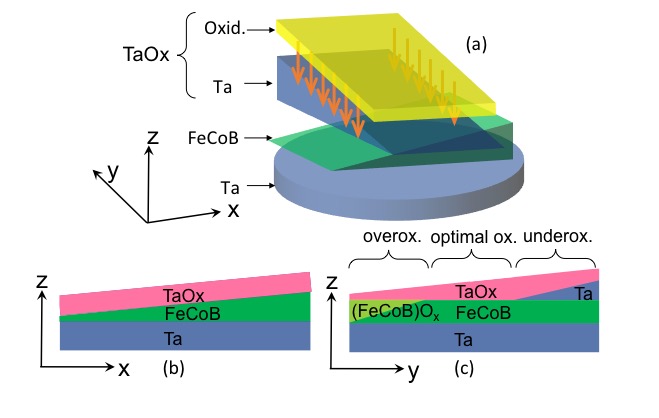}
	\caption{(a) Double wedge of Ta/FeCoB/TaOx/Al: a wedge of Ta deposited perpendicular to the wedge of FeCoB, followed by  oxidation. (b) View of the sample in the x-z plane for a given oxidation state of TaOx. (c) Along the y-axis for constant FeCoB thickness, overoxidized, optimally oxidized and underoxidized zones are obtained.}
	\label{sampledep}
\end{figure}
Our sample was deposited by magnetron sputtering, using on-axis configuration \textit{i.e.} substrate directly in front of the target (for fixed thicknesses) and off-axis configuration \textit{i.e.} substrate shifted with respect to the target center (for thickness gradients). On a $100mm$ diameter Si wafer substrate with 500nm-thick thermally oxidized SiO$_2$, first a uniform underlayer of Ta ($3nm$) is deposited. This is followed by deposition of FeCoB (composition $Fe_{72}Co_8B_{20}$) wedge ($0.59-1.69nm$) using off-axis deposition technique along x-direction. Perpendicular to this wedge, \textit{i.e.} along y-direction a deposition of Ta wedge ($0.45-1.04nm$) is carried out followed by natural oxidation step (oxygen pressure $150mbar$ for $10s$) as illustrated in Fig. \ref{sampledep}(a). The oxidation of the Ta wedge leads to underoxidized and overoxidized Ta zones on the opposite edges apart from the optimal oxidation in the center as shown in Fig. \ref{sampledep}(c). For surface protection of TaOx, a thin capping layer of Al ($0.5nm$) was deposited. These two wedges  perpendicular to each other provide access to all combinations of thicknesses of FeCoB and oxidation state of the FeCoB/TaOx interface in the ranges of interest. The sample was further annealed at $225^\circ C$ for 30 minutes to improve PMA. 

\begin{figure*}
\centering
\includegraphics[width=0.98\textwidth]{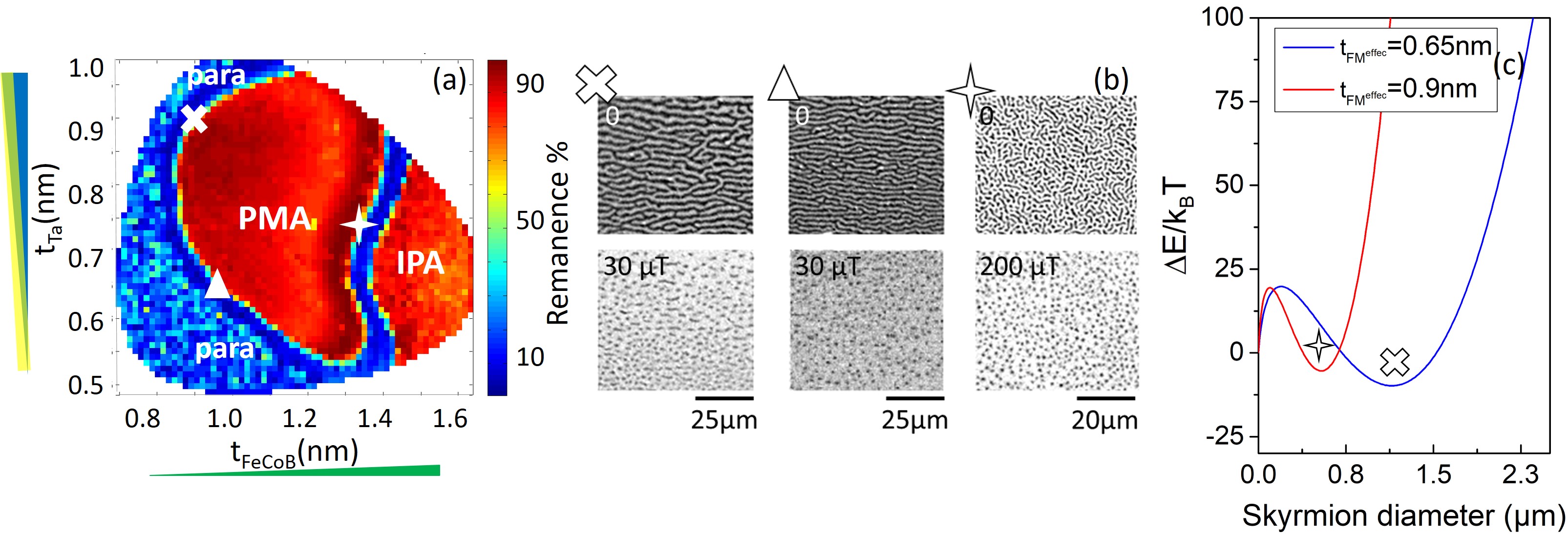}
\caption{(a) Remanence map of Ta/FeCoB/TaOx obtained by MOKE shows different regions: paramagnetic (para), PMA and IPA zones. Positions marked by cross, triangle and star represent the zones where the MOKE images of (b) were taken. (b) p-MOKE images under zero magnetic field (top) show stripes. With small perpendicular magnetic field  (bottom)  skyrmions are formed in the three regions. (c) Analytical model: variation of the energy difference between the isolated skyrmion and the saturated state as a function of the skyrmion diameter close to  PMA to paramagnetic transition (cross, $t_{FM^{effec}}=0.65nm$, $A=10pJ/m$, $M_s = 0.975 MA/m$, $D = 0.308 mJ/m^2$, $K_s = 0.419 mJ/m^2$, $\mu_0H=0.4737mT$) and to     PMA to IPA  transition (star, $t_{FM^{effec}}=0.9nm$, $A=10pJ/m$, $M_s = 0.985 MA/m$, $D = 0.44 mJ/m^2$, $K_s = 0.62 mJ/m^2$, $\mu_0H=1.37mT$). In both cases, a energy minimum for a skyrmion exists.}
	\label{remmap}
\end{figure*}

In order to map the magnetic properties along the double wedged sample, we used a commercial NanoMOKE3 wafer mapper by Durham Magneto-Optics Ltd, with a focused laser (steps of 1 mm). This magnetometer is based on Magneto-Optic Kerr Effect (MOKE) and due to its $45^\circ$ light incidence it is sensitive to both perpendicular and in-plane components of magnetization. Moreover, a 2D magnetic field  allows performing both perpendicular and in-plane hysteresis loops. With  our specific polarizer/analyzer settings and the 2D magnetic field, the shape of the hysteresis loop obtained at the measured point, reveals the magnetic easy axis. 
A map of the double wedge sample representing the remanence percentage (ie. remanent magnetization normalized to saturation magnetization) depending on FeCoB and top Ta thickness is shown in Fig. \ref{remmap}(a). The geometry of the off-axis deposition with respect to the circular target leads to uniform thickness deposited along circular lines. Therefore, plotting the remanence map as a function of thicknesses leads to a distorted map of the wafer.
The map reveals that the PMA region is limited to the thicknesses: $0.9-1.3nm$ of FeCoB, above which in-plane anisotropy is energetically favored due to an increase of the dipolar  contribution. 
Furthermore,  between the PMA and IPA region, we observe a transition with zero remanence. In this transition region, the effective anisotropy is close to zero as a result of the balance between surface anisotropy and shape anisotropy.  If we follow the FeCoB wedge along the $+x$ direction, we encounter paramagnetic zone, PMA and subsequently IPA as usually observed \cite{Jungblut, Johnson}.

For FeCoB thicknesses in the range $0.85 - 1.2 nm$, we also observe paramagnetic region along the two extremes of the Ta wedge that give zero remanence and noisy signal. For instance, if we follow the Ta wedge, corresponding to $t_{FeCoB} \approx 0.95 nm$, paramagnetic region is obtained for $t_{Ta} \textless 0.65 nm$ and $t_{Ta} \textgreater 0.92 nm$. For low Ta thickness, the excess oxygen, not trapped in TaOx layer, reaches FeCoB and partially oxidizes the ferromagnet. In this overoxidized region, the ferromagnetic layer thickness is thus decreased, leading to  lower  Curie temperature ($T_C$).  This is similar to the behavior observed in other systems such as Pt/Co/AlOx \cite{Marine}. More surprisingly, paramagnetic region is also observed for higher thicknesses of top Ta. This also corresponds to a decrease of Curie temperature \cite{fig1S}. In this case, we infer that TaOx is partially replaced by Ta at the top FeCoB interface (Fig.\ref{sampledep}(c), underoxidized region). Ta at the interface with FeCoB is known to result in dead layer (of the order of 0.3 nm \cite{Titiksha}), which is supported by ab-initio calculations \cite{Hallal}, showing a decrease of Fe magnetization. Similar results have been obtained in Ta/FeCoB/TaOx with an oxidation gradient of TaOx \cite{Yu2014}: on both sides of the optimum PMA for optimal oxidation, the hysteresis curves change was interpreted as a change in easy-axis. However, from their measurements, it could also be due to a transition towards paramagnetic phase.
To summarize, we can see that  the FeCoB/TaOx oxidation state largely affects magnetic order and in particular the Curie temperature that is lower both in the case of TaOx overoxidation or underoxidation.

Through this mapping we hence identified the different transition regions from PMA to paramagnetic for underoxidized and overoxidized cases and from PMA to IPA. In these transition regions, the magnetic order is weak (decrease of $M_s$ and $A$ in the transition from PMA to paramagnetic) and/or anisotropy is close to zero (in particular, the PMA to IPA transition is a signature of vanishing effective anisotropy); we thus expect low domain wall energy and easy nucleation of demagnetizing stripes and possibly skyrmions.  
We therefore performed polar-MOKE (p-MOKE) microscopy measurements to study the magnetic domain configuration and possible skyrmion nucleation sites. 

Here, we  study domains of a minimum size of 1 $\mu m$. The observed skyrmions of such sizes are often called skyrmionic bubbles, but share the same topology as small size skyrmions and we will thus call them skyrmions too. In the optimally oxidized TaOx zone, the PMA is high and large domains (hundreds of microns) are nucleated with magnetic field. The magnetization reversal on the application of out of plane magnetic field is dominated by domain wall propagation mechanism.
As we move from the PMA towards the paramagnetic transition zone, the size of the domains at zero field reduces to become stripes as shown in Fig. \ref{remmap}(b) top, cross and triangle corresponding to the positions marked on the remanence map (Fig. \ref{remmap}(a)). The stripe like domains are observed to be oriented along the frontier of the PMA to paramagnetic transition zone. Since the magnetic parameters like $M_s$ (thus $T_C$) and $K_s$ drop abruptly in a direction perpendicular to this transition zone, the stripes are elongated along the direction where the variation of these parameters is less pronounced. On the application of a small out-of-plane magnetic field ($30\mu T$), skyrmions are nucleated (Fig. \ref{remmap}(b) bottom). The Zeeman energy in this case helps balancing out the other interactions and the presence of DMI leads to a stable skyrmion phase. These skyrmions appear quite isolated and non interacting with each other. The value of DMI in this zone was measured by Brillouin Light Spectroscopy (BLS) and was found to be around $0.08mJ/m^2$ \cite{Titiksha}, which is consistent with the values in the literature for this system \cite{Jiang,Chaurasiya}. On further increasing the magnetic field, a uniformly magnetized state becomes favorable. Interestingly the oriented stripes and their evolution towards skyrmions under small magnetic field are obtained for all tested PMA to paramagnetic transition regions independent on the oxidation state of the TaOx.

Close to the other transition zone between PMA and IPA, a demagnetized domain structure prevails too at zero field. However, in this case the domain structure is isotropic labyrinthine like (Fig. \ref{remmap}(b) top). The FeCoB is thicker in this region and hence any overoxidized FeCoB layer or Ta/FeCoB mixed layer has a lesser relative contribution to the total energy by comparison with the PMA to paramagnetic transition region. The magnetization is nearly uniform in both the horizontal (along FeCoB wedge) and vertical (along TaOx wedge) directions and only effective magnetic anisotropy changes when FeCoB thickness is varied. This can be the reason that leads to isotropic labyrinthine-like demagnetized domain structure  as compared to the well-oriented stripes in the PMA to paramagnetic transition zone, as shown in Fig. \ref{remmap} (b). In this PMA to IPA transition region, application of a small out of plane magnetic field also stabilizes skyrmions. 

In order to better understand the formation of skyrmions is these very different regions of the sample, we used the analytical model of isolated skyrmion developed in \cite{Marine}. We calculated the skyrmion energy relative to the saturated state for the set of experimental parameters extracted at different transition regions. This model incorporates the contributions from dipolar, exchange, anisotropy, Zeeman and DMI energies as a function of the skyrmion diameter. 
The total energy is calculated and compared with the thermal energy ($k_BT$) to estimate the thermal activation barrier (Fig. \ref{remmap}(c)). The effective ferromagnetic thickness $t_{FM}^{effec}$ is taken into account, ie. we have removed the dead layer thickness from the nominal thickness of FeCoB  (which is the one represented in Fig. \ref{remmap} (a)).
By using two sets of experimentally extracted magnetic parameters, respectively corresponding to the PMA to paramagnetic transition on the one hand (cross or triangle regions, giving similar magnetic parameters) and to PMA to IPA transition on the other hand (star), we managed to show that a stable energy minimum  (ie. a skyrmion) exists for both cases, confirming the experimental observations of stable skyrmions close to these transition regions. When comparing PMA to IPA and PMA to PM transition regions, we obtain a slight decrease of magnetization (1\%) and strong decreases of both surface DMI $D_s=D\times t$ (30\%) and surface anisotropy (30\%). Both these surface quantities are linked to surface quality thus similar variations are not surprising.

\begin{figure}
	\centering
	\includegraphics[width=0.65\columnwidth]{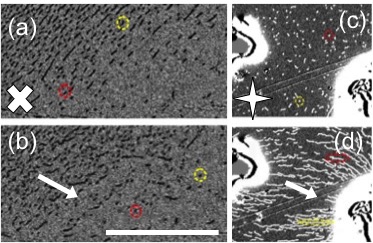}
	\caption{p-MOKE images of skyrmions under applied current  for two transition regions (cross (a-b) and star (c-d)). The scale, 50$\mu m$ corresponding to white bar in (b), is the same for all images. Images before (a) (resp. (c)) and after (b) (resp. (d)) 5 sec.  (resp. 8 sec.) of DC current application (direction of current given by white arrows).  (a,b) Skyrmions  move  uniformly, unpinned and undistorted.  (c,d) Most of the skyrmions get elongated and pinned along the direction of the current.  Red and yellow dotted circles trace the location of two skyrmions under current in the successive images.}
		\label{skyrmot}
\end{figure}
To qualitatively observe the motion of the skyrmions under applied current, we used conducting tips integrated with the p-MOKE setup. A perpendicular magnetic field was applied to stabilize the skyrmions while observing their motion under current. The skyrmions formed in the PMA to paramagnetic underoxidized transition zone  (cross region) move uniformly along the current flow i.e. against the electron flow, as shown in Fig. \ref{skyrmot}(a,b). The uniform motion of these skyrmions under applied current by spin-orbit torques \cite{Gambardella} confirms their non trivial topology and hence their skyrmionic nature and their homochirality. 
By contrast, the motion of the skyrmions formed in the PMA to IPA transition region (star region) seems to be dominated by pinning. It can be seen in Fig. \ref{skyrmot} (c,d) that on the application of current, the skyrmions elongate in the direction of current. The acquired video of the motion \cite{videos} reveals that a few skyrmions demonstrate successive elongation and contraction \cite{fig2S} and move forward in the same direction again indicating a homochiral nature. Some however remain pinned, some disappear and some expand continuously. More pinning and instabilities seem to be present in this part of the sample.


We now compare our different skyrmions  with other studies. 
In our previous studies, we studied skyrmions formed close to transition from PMA to paramagnetic region (Pt/Co/AlOx \cite{Marine}, Ta/FeCoB/TaOx  \cite{Titiksha}) or PMA to IPA region (Ta/FeCoB/TaOx  \cite{Je}). The type of transition in studies by other authors is less easy to determine as optimal oxidation conditions and dead layers might differ between different deposition tools. However, we may infer that in the case of Ta/FeCoB/Ta/MgO \cite{Yu2016}, ie. with an insertion of Ta, it will lead to a lower oxidation in the presence of more Ta, similar to our underoxidized region and that studies of Ta/FeCoB/TaOx \cite{Jiang} and Pt/CoFeB/MgO \cite{Boulle} with thicker ferromagnets correspond to PMA to IPA transitions. Let's notice that the stripes in these different studies have the same trends as in our case: they are more isotropic in the case of a PMA to IPA transition (\cite{Jiang, Boulle} and \cite{Yu2016} for low Ta insertion) and more oriented along a specific direction for the PMA to paramagnetic case (\cite{Marine, Titiksha} and \cite{Yu2016} in the case of thicker Ta thus more underoxidized interface).
The general trends are that skyrmions formed at the transition from PMA to paramagnetic region are generally bigger in size ($\mu m$), are thermally less stable (sometimes blinking), with very narrow location on the wafer but exhibit uniform current induced motion with negligible pinning. By contrast, the skyrmions nucleated at the PMA to IPA transition region are generally more stable, very often smaller in size (in the order of 100 nm), with wider location on the wafer, but are frequently pinned, partially hindering their current driven motion. However, micron scale skyrmions may be observed (this study and \cite{Yu2016}) close to  PMA to IPA transition, which means that the size is probably not specific to the transition type. At this transition, small skyrmions could probably be observed by moving towards thicker FeCoB with higher resolution methods.


These results suggest that skyrmions properties (size, time or field stability, current induced motion) depend on the type of transition region and the exact magnetic properties where they are nucleated. Although their properties could be tailored dynamically using a magnetic  or  electric field, the boundary conditions/ limit is dependent on the magnetic properties specific to the type of anisotropy transition. The different behaviors of these skyrmions under current may open distinct application objectives for them depending on the operating time scales.  Moreover, we anticipate that some trends observed in Ta/FeCoB/TaOx trilayer may be generalized to many HM/FM/MOx systems.


To conclude, we have shown that skyrmions  can be efficiently searched for in HM/FM/MOx trilayer structures by using crossed double wedges which allow to vary both the FM thickness and the oxidation state at the FM/MOx interface. A single 100 mm diameter wafer thus provides all combinations of magnetic parameters that form either PMA, IPA or paramagnetic regions. We have observed for HM=Ta, FM=FeCoB and MOx=TaOx, that skyrmion phases are obtained in the vicinity of transitions between these different magnetic phases. We also observed that these skyrmions may be moved by current and may present different current induced behaviour depending on the  transition region where they are nucleated. Our results highlight the fact that these transition zones are not equivalent and each of them might lead to a different applicative purpose. 

\begin{acknowledgements}
The authors acknowledge funding by the French ANR (contract no. ELECSPIN ANR-16-CE24-0018), by the
University Grenoble Alpes (project Automag2D), and by the Nanosciences Foundation. The authors thank Gilles Gaudin and André Thiaville for fruitful discussions.

\end{acknowledgements}

\end{document}